\documentclass[%
 reprint,
superscriptaddress,
 amsmath,amssymb,
 prl,
floatfix,
]{revtex4-2}

\usepackage{graphicx}
\usepackage{dcolumn}
\usepackage{bm}

\begin{document}

\preprint{APS/123-QED}

\title{Impact of the valley orbit coupling on exchange gate for spin qubits in silicon quantum dots}

\author{Bilal Tariq}
\email{bilaltar@buffalo.edu}
\affiliation{%
Department of Physics, University at Buffalo, SUNY, Buffalo, New York 14260, USA.
}%
\affiliation{%
National Center for Physics, Quaid-i-Azam University Campus, Islamabad 44000, Pakistan
}
\author{Xuedong Hu}%
 \email{xhu@buffalo.edu}
\affiliation{%
Department of Physics, University at Buffalo, SUNY, Buffalo, New York 14260, USA.
}%

\date{\today}

\begin{abstract}
The presence of degenerate conduction band valleys and how they are mixed by interfaces play critical roles in determining electron interaction and spectrum in a silicon nanostructure.  Here we investigate how the valley phases affect the exchange interaction in a symmetric two-electron silicon double quantum dot. Through a configuration interaction calculation, we find that exchange splitting is suppressed at a finite value of valley phase difference between the two dots, and reaches its minimum value ($\sim 0$) when the phase difference is $\pi$. Such a suppression can be explained using the Hubbard model, through the valley-phase-dependent dressing by the doubly occupied states on the ground singlet and triplet states.  The contributions of the higher orbital states also play a vital role in determining the value of the exchange energy in general, which is a crucial parameter for applications such as exchange gates for spin qubits.

\end{abstract}

\maketitle

\noindent {\it Introduction:} The spin of an electron or a nucleus confined in a semiconductor nanostructure is a qubit with intriguing potential for scalability \cite{loss1998quantum, kane1998silicon}.  Out of a multitude of material platforms and encoding schemes \cite{hanson2007spins, zwanenburg2013silicon, kane1998silicon, borselli2011pauli, divincenzo2000universal, petta2005coherent, shi2012fast, benito2020hybrid, hendrickx2020fast}, silicon is particularly enticing as a host for spin qubits because of its low abundance of spinful isotopes, which can be further reduced through isotopic enrichment.  As a result, electron spins have particularly long coherence times in Si \cite{veldhorst2014addressable, muhonen2014storing, borjans2019single}. 
Furthermore, exchange interaction, which originates from Coulomb interaction and Pauli principle, is inherently strong and allows fast two-spin gates \cite{burkard1999coupled, hu2001spin, burkard2002cancellation, koiller2001exchange, harvey2017coherent,  yang2017suppression}.  These favorable properties, together with ingenuity from experimentalists, have led to impressive achievements such as high fidelity single-qubit \cite{yoneda2018quantum, yang2019silicon} and two-qubit gates \cite{ sigillito2019coherent, veldhorst2015two, watson2018programmable, chan2021exchange}

A scalable qubit needs to be reproducible in its properties.  For electron spin qubits in Si, one of the main concerns has been the valley degree of freedom, i.e. the degeneracy in the Si conduction band, with focus on the valley-orbit coupling induced by the interface, and the associated effects \cite{koiller2001exchange, hada2004exchange, friesen2006magnetic, friesen2007valley, goswami2007controllable, shi2013spin, zwanenburg2013silicon, benito2019optimized, huang2021electric}. The study of this problem in single quantum dots have focused on the magnitude of valley splitting and interesting phenomena such as the spin-valley hotspot for spin relaxation \cite{yang2013spin, huang2014spin, borjans2019single, hollmann2020large}.  It has also been recognized in recent years that valley orbit coupling is generally complex, and its phase, particularly its variations across neighboring quantum dots, plays a crucial role in determining the tunnel coupling between dots \cite{burkard2016dispersive, gamble2016valley, zhao2018coherent, tagliaferri2018impact, tariq2019, ferdous2018valley, hosseinkhani2020electromagnetic, voisin2020valley, borjans2021probing}.

In this Letter we study how exchange coupling in a Si double quantum dot (DQD) is affected by valley physics, particularly valley-orbit coupling.  In earlier studies valley physics was often ignored under the assumption that valley splitting is large \cite{li2010exchange, culcer2010quantum, culcer2010interface, friesen2010theory, friesen2007valley} or the valley phase variation is small and can be treated perturbatively \cite{culcer2012valley}.  However, our recent study has shown that a single interface step could lead to an almost $\pi$ phase shift in the valley phase and/or a strong suppression of the valley splitting \cite{tariq2019}.  Therefore here we pay close attention to cases where valley splitting is small, or where the valley phase difference across a DQD is large.  We find that the two-electron exchange coupling in a symmetric Si DQD depend sensitively on the valley phase difference between the dots, and can be strongly suppressed even when valley splitting is large in both dots.  If valley splitting is small in at least one of the dots (we will define precisely what we mean by ``small''), the exchange gate protocol may have to be re-envisioned altogether because of the presence of additional singlet and triplet states that participate in the low-energy two-electron dynamics.  We have also explored the impact on the exchange splitting by the valley-orbit coupling to excited orbital states.  In short, our results clearly demonstrate the challenges posed by the valley-orbit physics on exchange gates for spin qubits in Si, and outline the necessary steps toward reliable exchange gates.

{\it{Theoretical Model: }}
We calculate the exchange splitting between the ground singlet and triplet states of a symmetric two-electron Si DQD using the configuration interaction (CI) approach.  We start with the simplest basis set, consisting of only the S-orbitals within each dot.  This is the equivalent to a Hund-Mullikan calculation but with two valley eigenstates from each dot, therefore including all the crucial ingredients.  We also extend to larger basis sets that include up to D-orbitals for each dot in order to validate our results.

The single-electron basis states underlying our two-electron calculations are orthonormalized single-dot envelope functions multiplied by valley eigenstates that contain the local valley-orbit phases, as discussed in the Supplementary Materials.

With the valley-orbit phase generally different in the two dots, an electron can tunnel between any pair of single dot states, characterized by the intra- and inter-valley tunnel coupling matrix elements [by ``intra'' we mean that states in both dots are in the ground (excited) valley eigenstates]: 
\begin{eqnarray}
\nonumber
    t_{++} = t_{--} = \frac{t_0}{2}\left(1+e^{-i\phi}\right),\; \;
    t_{+-} = t_{-+} = \frac{t_0}{2}\left(1-e^{-i\phi}\right)
\end{eqnarray}
Here $t_0$ is the tunnel coupling within the same bulk valleys ($z$ or $-z$).  $\phi = \phi_L - \phi_R$ is the valley phase difference in the double dot, with $\phi_L$ and $\phi_R$ the valley-orbit phases of the left and right dot, respectively. Without loss of generality, we choose $\phi_R=0$ and $\phi_L$ as a variable in our exchange energy calculations. In a physically realistic situation, both valley phase and valley splitting vary as functions of the interface roughness, such as the location of an interface step, and their variations may be correlated. However, to clarify their individual influences on the exchange splitting, we treat them as independent variables for most of this study.

An electric field along the axis of the DQD shift the energy levels of the two dots relative to each other, moving the DQD into the detuned regime. A particularly interesting parameter space is when the lowest double occupied singlet state is near resonance with the two-dot singlet, where exchange spitting is dominated by this coupling between the singlet states and is quite tunable \cite{petta2005coherent}. However, it is also well known that effect of charge noise is particularly strong in the detuned regime \cite{shulman2012demonstration}.  We thus restrict ourself to the zero detuning (symmetric) point, where the system is insensitive to the charge noise to the first order \cite{reed2016reduced, martins2016noise}.

{\it{Exchange Energy with S orbitals: }}
Including the valley degree of freedom, the minimal CI model to calculate the exchange splitting in a Si DQD is the  Hund-Mullikan model, which includes the ground orbital (S orbitals of the Fock Darwin states in the in-plane directions) in each valley in each of the two quantum dots.  From these four single-electron orbitals (after orthonormalization), one can form 10 symmetric and 6 anti-symmetric two-electron orbital states.

Specifically, labeling the S-orbitals in the two dots as $L$ and $R$, and the valleys as $+$ and $-$ (with valley splittings $\Delta_L$ and $\Delta_R$), we can form four two-dot symmetric or anti-symmetric states: $(L_-R_-,L_-R_+,L_+R_-,L_+R_+)$, two single-dot anti-symmetric double occupied states: $(L_-L_+,R_-R_+)$, and six single-dot symmetric double occupied states: $(L_-L_-,L_+L_+, L_-L_+, R_-R_-,R_+R_+,R_-R_+)$.  We then use these basis states to expand the two-electron Hamiltonian and obtain the singlet and triplet spectrum, respectively.

\begin{figure}[b]
\minipage{0.4\textwidth}
  \includegraphics[width=7.1cm, height= 8.5cm]{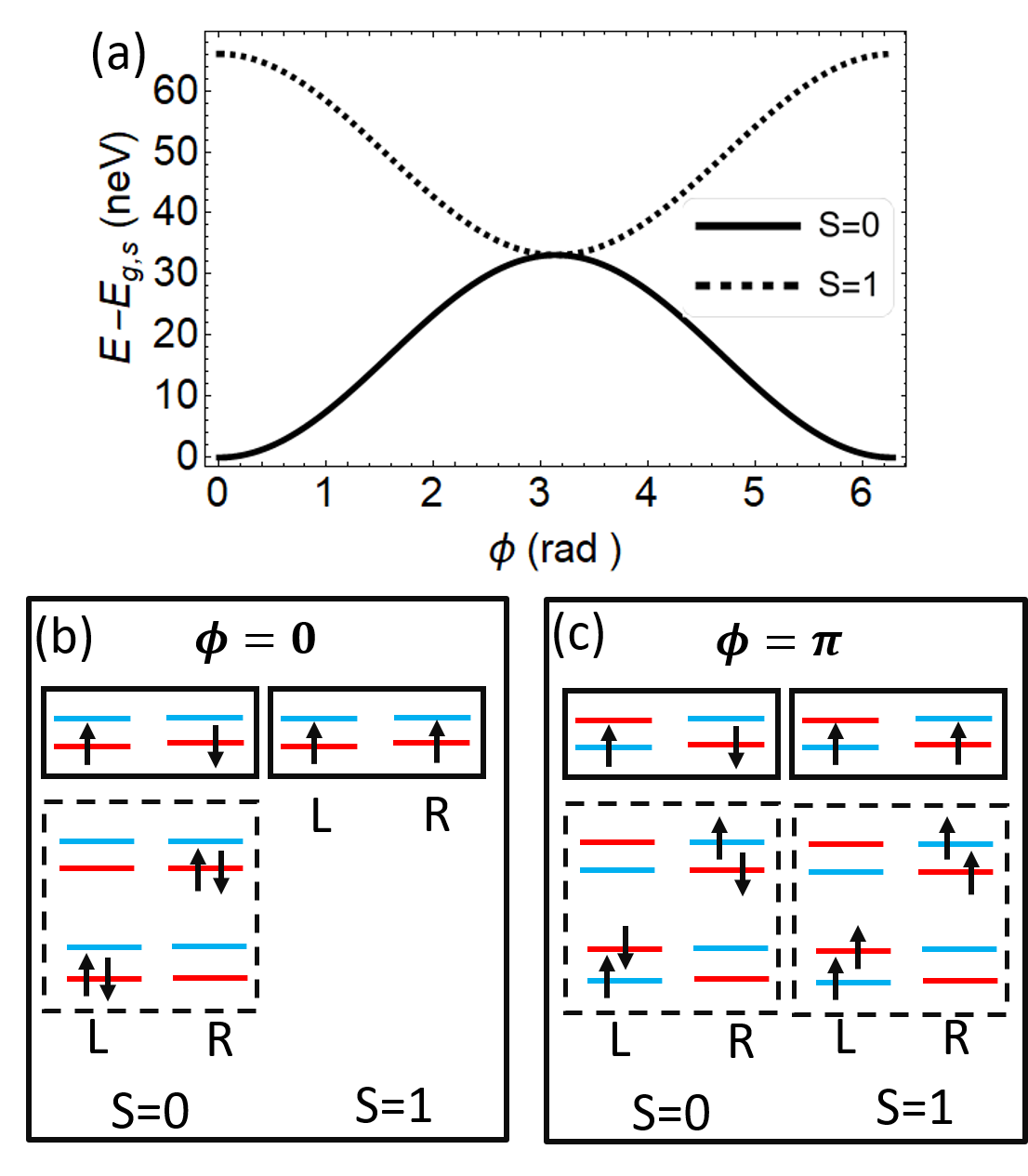}
\endminipage\hfill
   \caption{Panel (a): Dependence of the ground singlet (S$=$0) and triplet (S$=$1) energy levels on the valley phase difference between the two dots (we take $\phi_R=0$). For a smooth interface, the exchange interaction is $E_J=67$ neV with $\Delta_L=\Delta_R=0.1$ meV. Simple pictures of state dressing for phase difference $\phi=0$ and $\phi=\pi$ are shown in panels (b) and (c), respectively. We use the ground singlet energy level ($E_{g,s}$) for smooth interface as our reference. 
}\label{fig_1}
\end{figure}

In Fig.~\ref{fig_1}a we show the ground singlet (solid) and triplet (dotted) energies as functions of the interdot valley phase difference $\phi = \phi_L$ ($\phi_R = 0$ is fixed), with the valley splitting in both dots set at $0.1$ meV.  For a smooth interface, when $\phi_L= \phi_R = 0$, the value of the exchange splitting is at a maximum of 66 neV. This value depends on the tunnel coupling, quantum dot confinement, and coulomb interaction, and can be tuned easily by changing the height of the barrier potential.  In this example calculation the dot radius is set at 8 nm (the corresponding orbital excitation energy is 6.27 meV, and the onsite Coulomb energy is 16.1 meV) and the interdot distance at 40 nm, making sure that tunnel coupling is quite small.  With both quantum dots having the same valley phase, the electrons experience the so-called valley blockade: an electron in $L_{\pm}$ state can only tunnel to the $R_{\pm}$ state as they have the same underlying Bloch states, while tunneling between $L_\pm$ and $R_\mp$ are forbidden as their underlying Bloch states are orthogonal. The exchange splitting between the ground singlet and triplet states can thus be calculated within the $(L_-L_-,R_-R_-,L_-R_-)$ block of the block-diagonal Hamiltonian, and the additional valley states do not contribute to the ground singlet-triplet exchange splitting.  The situation is thus quite similar to the Hund-Mullikan model for a GaAs DQD \cite{burkard1999coupled, hu2000hilbert}.   

As the valley phase difference in the double dot increases from zero, valley blockade is lifted.  An electron can tunnel between any pair of valley states in the two dots.  In other words an electron in $L_{-}$ can tunnel to both $R_{-}$ and $R_{+}$, and vice versa.  Consequently, as shown in Fig.~\ref{fig_1}(a), the exchange splitting decreases, and eventually vanishes when the phase difference reaches $\pi$.  The physical picture can be most clearly illustrated by comparing panels (b) and (c) of Fig.~\ref{fig_1}.  In Fig.~\ref{fig_1}(b), where $\phi = \phi_L - \phi_R = 0$, the energy of the ground singlet state is lowered by the dressing from the doubly-occupied states $L_{-} L_{-}$ and $R_{-}R_{-}$, while the ground triplet is lowered by the Coulomb exchange but cannot be dressed by the doubly occupied states $L_{-}L_{+}$ and $R_{-}R_{+}$ as they are decoupled due to valley blockade between $L_\pm$ and $R_\mp$.  On the other hand, in Fig.~\ref{fig_1}(c), where $\phi = \pi$, both ground singlet and triplet states benefit from dressing by the doubly occupied states $L_-L_+$ and $R_-R_+$.  Doubly occupied states $L_-L_-$ and $R_-R_-$, which can only be singlet, do not couple to the ground singlet state because of the orthogonality of their underlying Bloch states.  As such the ground singlet and triplet states are dressed the same and their energies are lowered equally.  Furthermore, as discussed in the Supplementary Materials, the Coulomb exchange also has a dependence on the valley phase difference in the form of $\cos^2(\phi)$.  Including both influences, the exchange splitting decreases as $\phi$ increases from 0, and vanishes at $\phi = \pi$.

The results of Fig.~\ref{fig_1} are obtained with finite valley splittings of 0.1 meV in each dot.  They clearly show that valley phase difference between the two dots play a pivotal role in determining the exchange splitting even in the presence of finite valley splittings.  Furthermore, as we have demonstrated in Ref.~\onlinecite{tariq2019}, interface roughness in general affects both the phase and magnitude of valley-orbit coupling in a quantum dot.  Below we examine the influence of valley splittings on the exchange coupling by keeping the valley phases the same in the two dots. 

\begin{figure}[b]
\minipage{0.4\textwidth}
  \includegraphics[width= 7cm, height= 4.5cm]{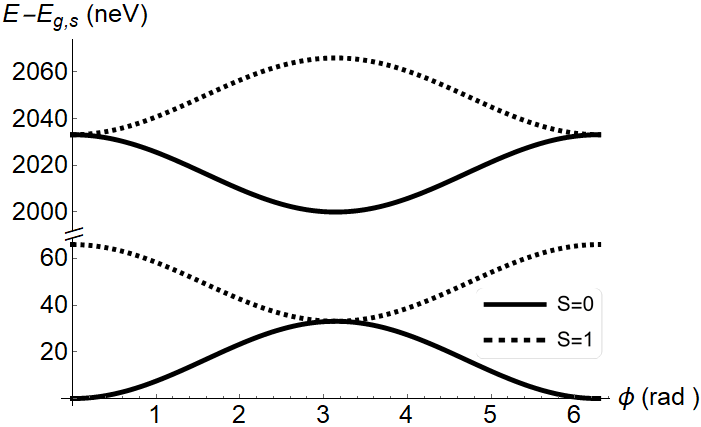}
\endminipage\hfill
   \caption{Energy levels of the lowest two singlet (S$=$0) and triplet states (S$=$1) as a function of $\phi$. We use $\Delta_R=1\mu eV$, $\Delta_L=100 \mu eV$, and $\phi_R=0$.
}\label{fig_2}
\end{figure}

{\it{Effects of valley splitting on exchange coupling: }} In a Si quantum dot valley splitting ranges from a few hundreds of $\mu$eV \cite{yang2013spin, gamble2016valley, hollmann2020large} to less than 10 $\mu$eV \cite{zajac2015reconfigurable, borjans2019single, burkard2016dispersive, borjans2021probing}.  
If the valley splitting is small, for example as compared with the thermal broadening of a nearby reservoir (at a typical electron temperature of 150 mK, the thermal broadening is about 10 $\mu$eV) for spin initialization, an electron could be initialized into the correct spin state but with a mixed valley state.  Such an unwanted orbital freedom may not affect single spin manipulation, assuming the two valley eigenstates having the same $g$-factor.  However, when considering two-spin exchange coupling, this additional freedom in valley occupation could lead to significant difficulties.

Consider a fictitious situation when the magnitude of the valley-orbit coupling in the right dot is two orders smaller than in the left dot: $\Delta_L=100\mu$eV and $\Delta_R=1\mu$eV.  Now the first excited singlet and triplet states are roughly 2 $\mu$eV above the ground singlet and triplet states, respectively, with the electron in the right dot occupying the excited valley state, i.e. $L_-R_+$.  In Fig.~\ref{fig_2} we plot the energies of the ground and first excited singlet and triplet states as functions of the valley phase difference. The ground states have the same behavior as in Fig.~\ref{fig_1}, while the phase-dependence of the excited singlet and triplet states are quite different. As such, the singlet-triplet splitting for the ground pair and the first excited pair are generally different.  If an electron is initialized into the right dot in a state $\left(\alpha |R_-\rangle + \beta |R_+\rangle \right)|\uparrow\rangle$, when tunnel coupling is allowed, the respective singlet-triplet states that can be formed from $|R_-\rangle$ and $|R_+\rangle$, i.e. the ground and first excited singlet-triplet pairs we plot in Fig.~\ref{fig_2}, will in general have different singlet-triplet splittings.  Consequently, the phase accumulated during an exchange gate will be different in these two pairs, making a spin swap gate almost impossible \cite{loss1998quantum}.

In short, a necessary condition for exchange gate protocol to be valid in a Si DQD is that the valley splitting in each dot is much larger than the thermal broadening of the reservoir used for initialization.  This condition guarantees a high fidelity preparation for a spin qubit in the ground valley eigenstate, taking away any uncertainty in the follow-up spin manipulations.

When $|\Delta_R|$ (and/or $|\Delta_L|$) is further reduced, the ground and excited singlet and triplet states for the DQD become even more compact in the energy spectrum, and their dynamics cannot be straightforwardly disentangled, as we discuss in the Supplementary Materials. While the physics at this limit is subtle and interesting, the DQD does not have any utility for spin qubit manipulation anymore as the system cannot be properly initialized and controlled.

{\it{Exchange Energy in the Presence of an Interface Step: }} In the model calculations above we vary either the valley phase difference or the valley splittings in a quantum dot as an independent variable.  In a realistic situation, however, both the magnitude and phase of the valley orbit coupling depend on interface roughness and interface electric field. As such they tend to change in a correlated manner, as has been illustrated in Si/SiGe heterostructures with a single atomic layer step at the interface inside a quantum dot \cite{tariq2019, gamble2013disorder, gamble2016valley, ferdous2018valley, friesen2007valley, doi:10.1021/acs.nanolett.7b01677}. Here we explore how exchange coupling in a DQD is affected by the presence of an interface step.  In particular, in a full CI calculation that includes the valley degree of freedom, we need to calculate the valley-orbit matrix elements among all the single-electron orbitals, which would allow us to better clarify the effects of the higher orbital states and the valley orbit coupling parameters on the ground state exchange splitting. 

Including the s-, p-, and d-orbitals for the in-plane wave function in each dot, the different valley orbit coupling terms can be summarized in a matrix as
\begin{equation}
\nonumber
\Delta=
\begin{pmatrix}
\Delta_{ss} & \Delta_{sp_x} & 0 & \Delta_{sd_{xx}} & 0 & 0\\
\Delta_{p_xs} & \Delta_{p_xp_x} & 0 & \Delta_{p_xd_{xx}} & 0 & 0\\
0 & 0 & \Delta_{p_yp_y} & 0 & 0 & 0\\
\Delta_{d_{xx}s} & \Delta_{d_{xx}p_x} & 0 & \Delta_{d_{xx}d_{xx}} & 0 & 0\\
0 & 0 & 0 & 0 & \Delta_{d_{xy}d_{xy}}  & 0\\
0 & 0 & 0 & 0 & 0 & \Delta_{d_{yy}d_{yy}} \\
\end{pmatrix} \,.
\end{equation}
Each of the term in $\Delta$ vary differently with the step position as shown in the Supplementary Materials Section S2. Given a step position and orientation, each of the matrix elements can be calculated straightforwardly.  The $\Delta$ matrix can then be included when calculating the orthonormal single-electron eigenbasis, over which we construct the two-electron states and calculate the exchange splitting.

\begin{figure}[t]
\minipage{0.4\textwidth}
  \includegraphics[width=7.1cm, height= 4.8cm]{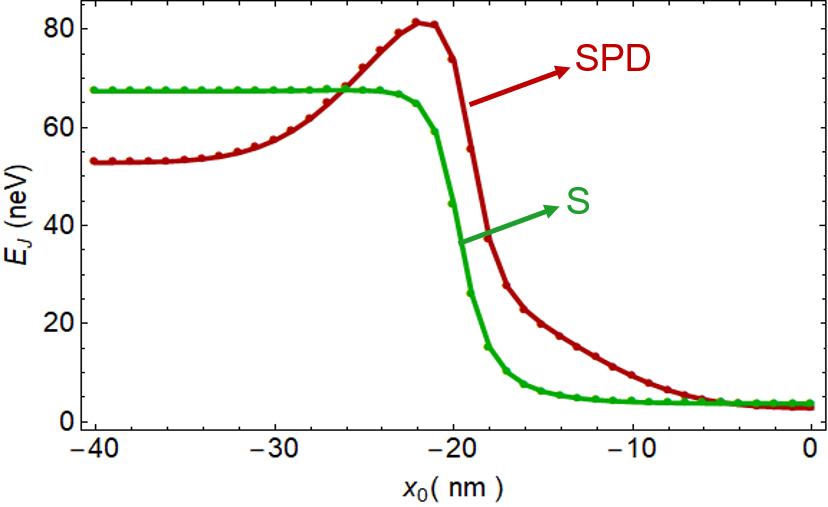}
\endminipage\hfill
   \caption{Exchange interaction as a function of the interface atomic step position, with different sizes of the basis set ($s$ only, or with $spd$ orbital states). The center of the DQD lies at the origin $x_0 = 0$. The center of the left dot is at $x_0=-20$ nm, and the dot radius is $\ell_0=8$ nm }\label{EX_spd}
\end{figure}

In Fig.~\ref{EX_spd} we plot the exchange splitting as a function of the step location $x_0$, with the step oriented perpendicular to the interdot axis.  $x_0 = 0$ refers to the situation when the step is at the midpoint between the two dots, and $x_0 = -20$ nm is when the step passes through the middle of the left dot. The most important feature in this figure is the suppression of exchange coupling when the step is located in between the two dots, similar to what we find in Fig.~\ref{fig_1}(a) when we only consider the s-orbitals.  This suppression has the same origin as well: when the step is in between the dots (at or near $x_0 = 0$), the valley phase difference between the dots is $\sim 0.85 \pi$, making the tunnel coupling between the ground valley states in the two dots very small and the exchange coupling strongly suppressed.

There are two additional features in the results of the exchange splitting shown in Fig.~\ref{EX_spd} when we include higher orbitals in our calculation. There is a broad peak when the step is in the middle of the left dot, and there is a longer tail of finite exchange splitting (as compared to the s-orbital only calculation) as the step approaches the middle between the dots. These features are mainly the results of a competition between two influences: the phase of the $\Delta_{ss}$, and the magnitude of $\Delta_{sp_x}$.  As discussed for Fig.~\ref{fig_1}, a non-vanishing phase for $\Delta_{ss}$ suppresses the magnitude of the exchange splitting for the ground singlet-triplet pair. On the other hand, the term $\Delta_{sp_x}$ originates from the symmetry breaking within the left dot due to the presence of the step. It reaches its maximum magnitude when the step is at the center of the dot, and causes a linear change in the exchange splitting as shown in the Supplement Material.  The valley-orbit coupling in the excited states plays an important role here because we have two small quantum dots, such that orbital excitation energy ($\sim 6.3$ meV) is much smaller than the onsite Coulomb interaction $\sim 16$ meV, making the dressing of the ground singlet and triplet states by the orbital excited states as important as the doubly occupied ground orbital states.  In other words, to achieve a numerical convergence for the exchange calculation, more orbital states need to be included.  However, for the purpose of exploring the qualitative effects of the valley-orbit coupling, our finite-size calculation here is sufficient.

{\it{Conclusion: }} In conclusion, we have performed analytical and numerical analysis of the ground singlet-triplet exchange splitting of a Si double quantum dot. Our results show that valley-orbit coupling in the two dots play crucial roles in determining the exchange energy.  In particular, it depends sensitively on the valley phase difference between the two dots, reaching a minimum when the phase difference is $\pi$, even in the presence of large valley splittings in both dots. We also show that it is imperative that valley splitting in each of the quantum dots should be large compared to the thermal broadening of the reservoir, such that a spin qubit can be properly initialized.  By examining the splitting in both the ground and first excited singlet-triplet pairs, we show that exchange gate would not work properly if both of these manifolds are involved in the spin dynamics.  Lastly, we show that the higher-energy orbital states also make important contributions in determining the value of the exchange energy, particularly for smaller dots with large on-site Coulomb interaction.

While our results are particularly relevant for Si/SiGe quantum dots, the phase dependence by the exchange coupling, irrespective of the magnitude of the valley splitting, is an important observation for SiMOS quantum dots as well, which tend to have larger valley splittings but also have an amorphous interface.  Our results shine a further spotlight on the interface roughness, and the need to understand and characterize them in order to achieve scalable quantum computing based on spin qubits in silicon.

{\it{Acknowledgments: }}
This work is partially supported by US ARO through grant W911NF1710257.

\newpage
\newpage

\begin{widetext}

\section{\large{Supplementary Materials}}

\section{ Tunnel Coupling of an Electron in a Double quantum dot}

Single-electron tunnel coupling is a crucial parameter in describing electron dynamics in a double quantum dot, and is an important matrix element for calculating exchange splitting \cite{li2010exchange, jiang2013coulomb}.  Here we describe how we calculate this quantity.  The Hamiltonian of an electron confined in a gate-defined silicon double quantum dot (DQD) is given by
\begin{equation}
    H(\boldsymbol{r})=-\frac{\hbar^2}{2m_t^2}\frac{\partial^2}{\partial x^2}-\frac{\hbar^2}{2m_t^2}\frac{\partial^2}{\partial y^2}-\frac{\hbar^2}{2m_l^2}\frac{\partial^2}{\partial z^2}+V_{\text{DQD}}(x,y)+V(z) +eEx
\end{equation}
where $m_t=0.192m_0$ and $m_l=0.98m_0$ are the transverse and longitudinal effective masses of an electron in each of the bulk conduction band valleys of silicon. $E$ is an electric field applied along the $x$ direction (axial direction of the double dot) used to provide an energy detuning between the left and right dot. $V(z)$ is a triangular confinement potential along the growth direction $z$ from the heterostructure barrier and the interface electric field. For a smooth interface, 
\begin{equation}
    V(z)=eF\left(z-z_I\right)+U_0\;\theta\left(-z+z_I\right)
\end{equation}
where $F$ is the magnitude of the interface electric field, and $U_0$ is the barrier potential. We assume the position of the interface at $z=z_I$. For a Si/SiGe hetrostructure we use $F=15$MV/m and $U_0=150$meV \cite{culcer2012valley} to obtain a $\sim 100 \mu$eV valley splitting, a typical value in the literature. For an interface with step we apply the same variational approach as previously used in Ref.~\cite{tariq2019}.

We model the in-plane confinement potential of the double quantum dot as a double quadratic well,
\begin{equation}
    V_\text{{DQD}}=\frac{\hbar^2}{2 m_l \ell_0^4}\text{Min}\left[(x+d)^2+y^2,(x-d)^2+y^2\right] \,,
    \label{VB_DQD}
\end{equation}
where $\ell_0$ is the characteristic length (radius of the ground Gaussian wave function) of each circular quantum dot and $2d$ is the distance between the centers of the two dots along the $x$ direction. The height of the barrier potential is controlled by changing the quantum dot size or the separation between the dots. In this paper, we have fixed $\ell_0=8$ nm ($E_s=6.3$ meV) and $d=20$ nm in our calculations. The height of barrier potential between the two dots is thus $19$ meV, such that there are three energy levels on each dot [i.e. the $s$, $p$, and $d$ two-dimensional Harmonic Oscillator (or Fock-Darwin) states] below the barrier potential at zero magnetic field.   

We adopt the envelope function approach within the effective mass approximation.  The envelope function of an electron in the left (right) dot is $L({\bf r})$  ($R({\bf r})$) without including the valley degree of freedom,
\begin{equation}
    L(\boldsymbol{r})=F(x+d,y)\phi(z),\;\;\; R(\boldsymbol{r})=F(x-d,y)\phi(z) \,,
\end{equation}
where $F(x\pm d,y)$ is the shifted Fock Darwin state in-plane, and $\phi(z)$ is the modified Fang-Howard function along the $z$ direction. The overlap of the wave function between the two dots is non zero, i.e. $\langle L(\boldsymbol{r})|R(\boldsymbol{r})\rangle \neq 0$. We thus first normalize the wave function as, 
\begin{equation}
    \begin{pmatrix}
    \bar{L}(\boldsymbol{r}) \\
    \bar{R}(\boldsymbol{r}) \\
    \end{pmatrix}=
   \begin{pmatrix}
    \langle L(\boldsymbol{r}) |L(\boldsymbol{r})\rangle & \langle L(\boldsymbol{r}) |R(\boldsymbol{r})\rangle\\
    \langle L(\boldsymbol{r}) |R(\boldsymbol{r})\rangle & \langle R(\boldsymbol{r}) |R(\boldsymbol{r})\rangle\\
    \end{pmatrix}^{-\frac{1}{2}}
    \begin{pmatrix}
    L(\boldsymbol{r}) \\
    R(\boldsymbol{r}) \\
    \end{pmatrix} \,,
\end{equation}
where $\{\bar{L}(\boldsymbol{r}), \bar{R}(\boldsymbol{r}) \}$ have been orthonormalized.  With only $S$ orbital state in each dot, the procedure can be done analytically. When higher orbital states are included, we calculate the orthonormalization coefficients numerically. 

In the presence of valleys, the electron wave function at a conduction band minimum can be written as a product of the envelope function and the underlying Bloch states,
\begin{eqnarray}
    \bar{L}_z(\boldsymbol{r})&=&\bar{L}(\boldsymbol{r})\; u_z(\boldsymbol{r})e^{-ik_0z},\;\;\;  \bar{L}_{-z}(\boldsymbol{r})=\bar{L}(\boldsymbol{r})\; u_{-z}(r)e^{ik_0z}\\
    \bar{R}_z(\boldsymbol{r})&=&\bar{R}(\boldsymbol{r})\; u_z(\boldsymbol{r})e^{-ik_0z},\;\;\;  \bar{R}_{-z}(\boldsymbol{r})=\bar{R}(\boldsymbol{r})\; u_{-z}(r)e^{ik_0z} \,,
\end{eqnarray}
where $u_z(\boldsymbol{r})$ and $u_{-z}(\boldsymbol{r})$ are the periodic parts of the underlying Bloch states and $\pm k_0 = \pm0.085\times 2\pi /a_{Si}$ represents the $z$ and $-z$ band minima of Si within the First Brillouin Zone, with $a_{Si}=0.543$ nm. Note that here $\langle \bar{L}_z|\bar{L}_{-z}\rangle=0$ due to the orthogonality of Bloch states. The one-electron Hamiltonian within these single-bulk-valley eigenbasis take the form,
\begin{equation}
    H=
   \begin{pmatrix}
    \bar{E}_L-\epsilon & \bar{\Delta}_L & \bar{t}_0 & 0\\
    \bar{\Delta}^*_L & \bar{E}_L-\epsilon  & 0 & \bar{t}_0\\
      \bar{t}_0 & 0 &\bar{E}_R+\epsilon & \bar{\Delta}_R \\
      0 & \bar{t}_0 & \bar{\Delta}^*_R & \bar{E}_R+\epsilon  \\
    \end{pmatrix} \,.
\end{equation}
The parameters here, including their definitions, are given in Table \ref{tab1}.
\begin{center}
\setlength{\tabcolsep}{18pt}
\renewcommand{\arraystretch}{1.2}
\begin{table}[h]
    \centering
    \begin{tabular}{ |p{3.5cm}|p{4.5cm}|p{2.5cm}|  }
  \hline
 Ground Orbital Energy & \;\;  $\bar{E}_{L}=\langle \bar{L}_{ \pm z}(\boldsymbol{r})|H_0|\bar{L}_{\pm z}(\boldsymbol{r})\rangle$   & \;\;\;6.3\;meV \\
  & \;\; $\bar{E}_{R}=\langle \bar{R}_{\pm z}(\boldsymbol{r})|H_0|\bar{R}_{\pm z}(\boldsymbol{r})\rangle$   & \;\;\;6.3\;meV \\
  \hline
 Tunnel coupling & \;\;\;\;\;$\bar{t}_0=\langle \bar{L}_z(\boldsymbol{r})|H_0|\bar{R}_z(\boldsymbol{r})\rangle$  &$-17.1$\;$\mu$eV  \\
 \hline
 Valley splitting & $|\bar{\Delta}_L|=|\langle \bar{L}_z(\boldsymbol{r})|H_0|\bar{L}_{-z}(\boldsymbol{r})\rangle|$   & \;\;$100\;\mu$eV\\
 & $|\bar{\Delta}_R|=|\langle \bar{R}_z(\boldsymbol{r})|H_0|\bar{R}_{-z}(\boldsymbol{r})\rangle|$   & \;\;$100\;\mu$eV\\
 \hline
 Detuning Energy &\;\; $\epsilon_L=\langle \bar{L}_{\pm z} (\boldsymbol{r})|eFx|\bar{L}_{\pm z}(\boldsymbol{r})\rangle$   & \;\;$-\epsilon$ \\
 &\;\; $\epsilon_R=\langle \bar{R}_{\pm z} (\boldsymbol{r})|eFx|\bar{R}_{\pm z}(\boldsymbol{r})\rangle$   & \;\;$\;\;\epsilon$ \\
  \hline
\end{tabular}
    \caption{Energy parameters of an electron in the double quantum dot}
    \label{tab1}
\end{table}
\end{center}

%
%

In Table \ref{tab1} $\bar{\Delta}_L$ and $\bar{\Delta}_R$ are the valley-orbit coupling of the left and the right dot. These are in general complex quantities,
\begin{equation}
    \bar{\Delta}_L=|\bar{\Delta}_L|e^{-i\bar{\phi}_L},\;\;\; \bar{\Delta}_R=|\bar{\Delta}_R|e^{-i\bar{\phi}_R} \,,
\end{equation}
where $2|\bar{\Delta}_L|$ and $2|\bar{\Delta}_R|$ are valley splittings in the two dots and $\bar{\phi}_L$ and $\bar{\phi}_R$ are the corresponding valley phases of the left and right dots. For a smooth interface, the valley splittings and phases of the left and right dot are the same. 

With all the calculations in the Supplementary Materials and main text based on the orthonormalized basis, we remove the bar sign on each of the terms without loss of generality.

A typical value of the valley splitting in a Si/SiGe Hetrostructure is 0.1 meV, which can be further tuned by an applied electric field along the growth direction \cite{yang2013spin, hosseinkhani2020electromagnetic}. However, the external electric field has only a small effect on the valley phase as it is mostly determined by the location and quality of the interface \cite{saraiva2009physical, zimm}. For example, the steps present at the Si-SiGe interface affects the relative phase between the left and right dot strongly, ranging from $0$ to $\pi$ depending on the atomistic details of a step \cite{friesen2007valley, zwanenburg2013silicon}. Furthermore, interface steps also affect the magnitude of valley splittings. In most of our model calculations we treat valley splitting and valley phase as phenomenological variables and vary them independently, while in reality any interface roughness would affect both the magnitude and phase of the valley-orbit coupling simultaneously.  Thus we also examine the overall effect of a single interface step on the energy levels and exchange energy by changing its positions at the interface. 

\begin{figure}[h]
\minipage{0.85\textwidth}
  \includegraphics[width= 13cm, height= 4.6cm]{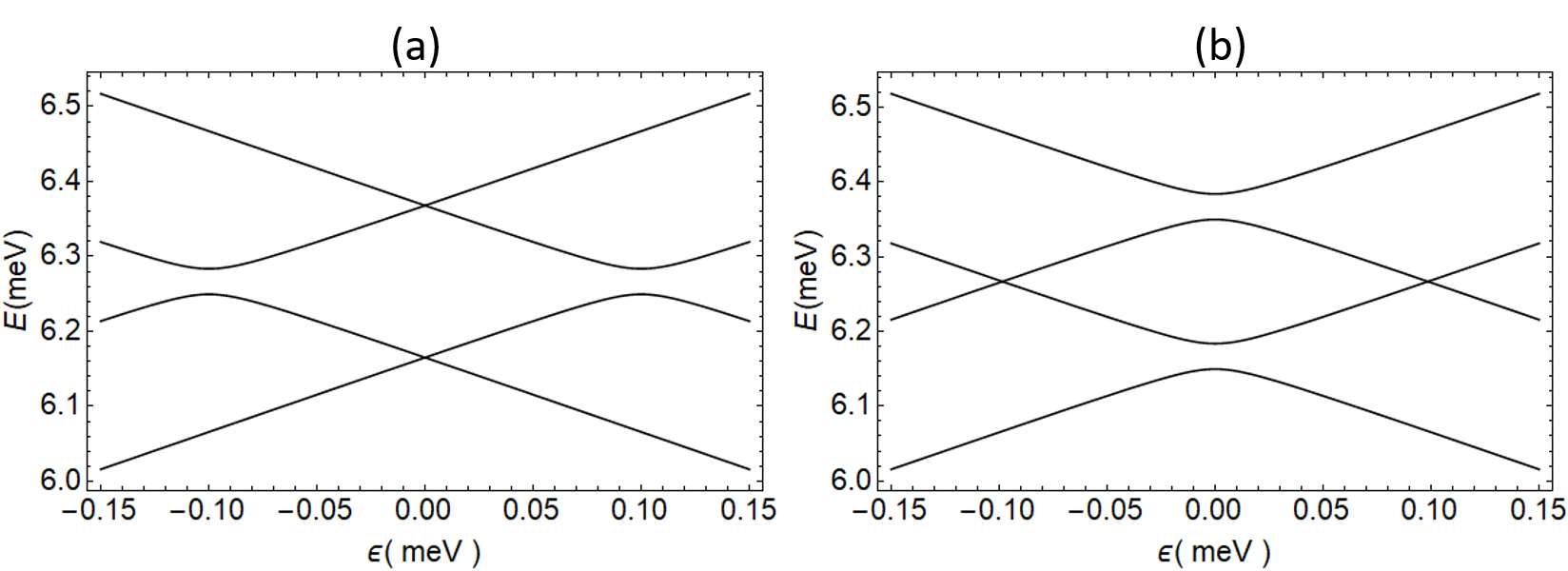}
\endminipage\hfill
   \caption{In (a) and (b) we have shown the energy levels of an electron of as a function of detuning energy at $\phi=0$ and $\phi=\pi$, respectively.
}\label{fig_s1}
\end{figure}

In Fig.~\ref{fig_s1} we show two examples of the energy spectrum of a single electron in a DQD as a function of the inter-dot detuning, when the phase difference between the dot is set at zero and $\pi$. Clearly, the energy levels in the DQD has a strong dependence on the valley phase difference between them as we observe a finite gap at zero detuning energy transformed into a levels crossing when the phase changes from 0 to $\pi$.
These results can be explained by using the single-dot valley eigen-basis $\{L_-,L_+,R_-,R_+\}$.  In terms of the bulk valley eigen-basis in the left and right dot, $\{L_z,L_{-z},R_z,R_{-z}\}$, the single-dot valley eigen-basis states are defined as,
\begin{equation}
    \begin{pmatrix}
    L_-(\boldsymbol{r})\\
    L_+(\boldsymbol{r})\\
      R_-(\boldsymbol{r}) \\
      R_+ (\boldsymbol{r}) \\
 \end{pmatrix}=
   \frac{1}{\sqrt{2}} \begin{pmatrix}
    1 & -e^{-i\phi_L} & 0 & 0\\
    1 & e^{-i\phi_L}  & 0 & 0\\
      0 & 0 &1 & -e^{-i\phi_R} \\
      0 & 0 & 1 & e^{-i\phi_R} \\
    \end{pmatrix}
    \begin{pmatrix}
    L_{-z}(\boldsymbol{r})\\
    L_{+z}(\boldsymbol{r})\\
      R_{-z}(\boldsymbol{r}) \\
      R_{+z} (\boldsymbol{r}) \\
   \end{pmatrix}.
\end{equation}
Accordingly, the single-electron Hamiltonian expressed in the single-dot valley eigen-basis is
\begin{equation}
    H=
   \begin{pmatrix}
    E_L-|\Delta_L|-\epsilon & 0 & t_{--} & t_{-+}\\
    0 & E_L+|\Delta_L|-\epsilon  & t_{+-} & t_{++}\\
      t^*_{--} & t^*_{-+} &E_R-|\Delta_R|+\epsilon & 0 \\
      t^*_{+-} & t^*_{++} & 0 & E_R+|\Delta_R|+\epsilon  \\
    \end{pmatrix} \,.
\end{equation}
The tunnel coupling matrix elements here depend on the valley phase difference between the two dots, 
\begin{eqnarray}
     t_1\equiv\bar{t}_{--(++)} &=&\langle \bar{L}_{-(+)}|H|\bar{L}_{-(+)}\rangle =\frac{t_0}{2}\left(1+e^{-i\phi}\right),\\
     t_2\equiv\bar{t}_{-+(+-)} &=&\langle \bar{L}_{-(+)}|H|\bar{L}_{+(-)}\rangle =\frac{t_0}{2}\left(1-e^{-i\phi}\right).
\end{eqnarray}\label{T1T2}
Here $-$ and $+$ subscripts refer to the lower- and higher-energy valley eigenstates, thus we refer to $t_{--}$ and $t_{++}$ as intra-valley tunneling while $t_{-+}$ and $t_{+-}$ as inter-valley tunneling coupling, with ``valley'' here implying single-dot valley eigenstates in each dot. In Fig.~\ref{fig_s2} we show the results of the electron tunneling results. It shows an electron tunneling within the same valley when $\phi_L=\phi_R=0$. This causes a finite energy gap at zero detuning region. If there is phase difference of $\pi$ between the left and the right dot, an electron can only tunnel from ground valley eigenstate in one dot to the excited state in the other, or vice versa. Hence at zero detuning, when the ground valley eigenstates in the two dots are resonant, we have a crossing between the energy levels. 

\begin{figure}[h]
\minipage{0.85\textwidth}
  \includegraphics[width= 7.5cm, height= 4.5cm]{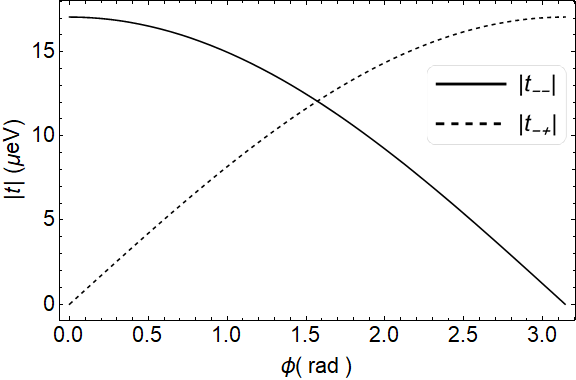}
\endminipage\hfill
   \caption{We show the tunnel coupling results within the same and different valley states from left to right dot as a function of valley phase.
}\label{fig_s2}
\end{figure}

The use of the valley eigenbasis has the advantage in providing a clear physical picture of the electron tunneling from one dot to the other. However, when valley splitting in either one or both of the dots are zero, valley phase becomes ill defined. Hence we go back to the left and right dot bulk valley basis set when we discuss the limit of zero valley-orbit couplings.

\begin{figure}[h]
\minipage{0.85\textwidth}
  \includegraphics[width= 13cm, height= 5cm]{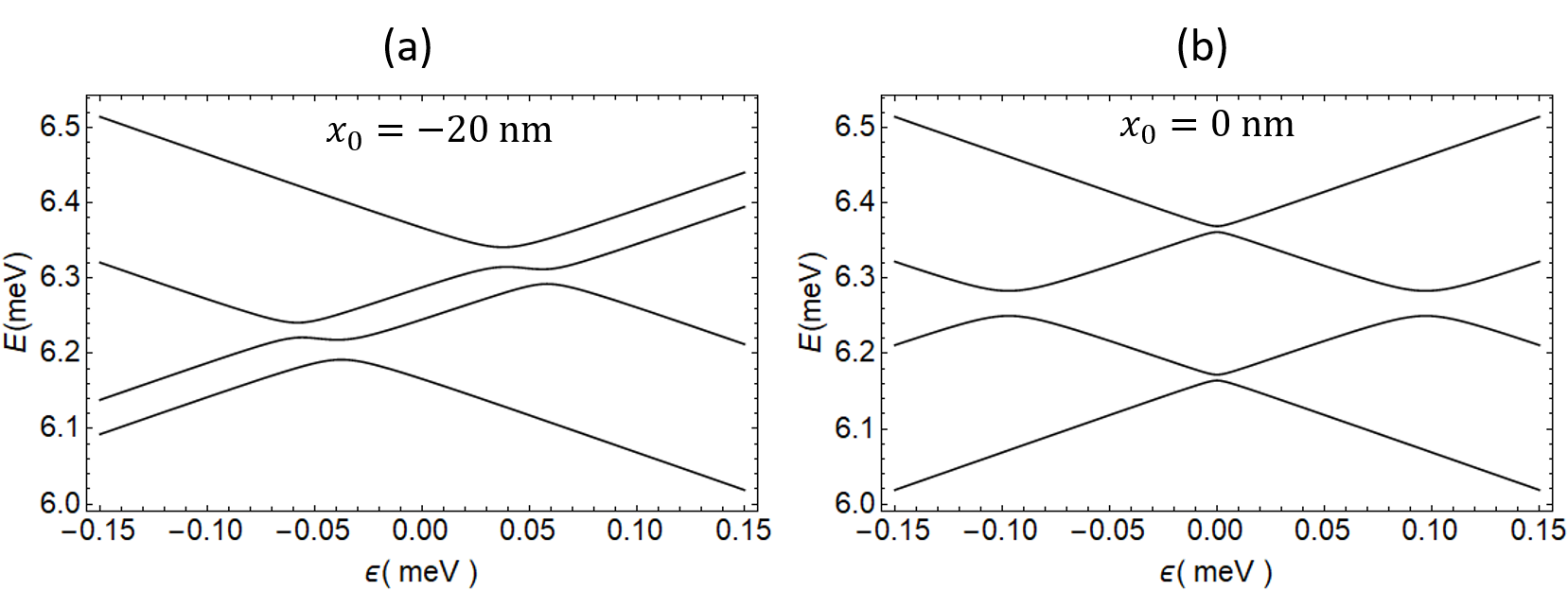}
\endminipage\hfill
   \caption{Panels (a) and (b) are energy levels of an electron in the presence of an interface step at position $x_0=-20$ nm and $x_0=0$ nm, respectively.
}\label{fig_s3}
\end{figure}

In Fig.~\ref{fig_s3}(a) and (b) We show two examples of energy spectrum of a single electron in a double dot in the presence of an interface step, with the step position at the center of the left dot and in the middle between the two dots, respectively. When the step is at $x=-20$ nm, the valley splittings and phases on the left and the right dots are quite different. The valley splittings in the left dot is reduced by 77\%., while the valley phase difference is $73^\circ$.  As such the spectrum in panel (a) is quite asymmetric, though all the level crossings become anti-crossings because both $t_{--}$ and $t_{+-}$ are finite.  On the other hand, when the step is at the middle between the two dots, the phases in left and right dot differ by $0.85 \pi$, while the valley splittings are equal.  We thus find a spectrum that is symmetric with respect to detuning, but with near crossings at zero detuning (from $t_{--} = t_{++}$) and large anticrossings at finite detuning (from $t_{-+} = t_{+-}$).

\section{Exchange Interaction of the two Electrons in a Double Quantum Dot}\label{S2}

Coulomb interaction between two electrons leads to a difference in the energy of the parallel and antiparallel spin orientations of the electrons due to the Pauli Exclusion principle \cite{burkard1999coupled}. This energy difference due to spin configuration is called exchange energy. Here we show in detail a minimal calculation to evaluate the exchange splittings for two electrons in a Si DQD. We include the S-orbital and two valley states in each quantum dot. The Hamiltonian for two electrons in a biquadratic DQD is defined as,
\begin{equation}
    H(\boldsymbol{r}_1,\boldsymbol{r}_2)=H(\boldsymbol{r}_1)+H(\boldsymbol{r}_2)+H_C(|\boldsymbol{r}_1-\boldsymbol{r}_2|),
\end{equation}
where $H(\boldsymbol{r}_1)$ and $H(\boldsymbol{r}_2)$ are the single electron Hamiltonian with $\boldsymbol{r}_1$ and $\boldsymbol{r}_2$ are the positions of the electrons. The Coulomb interaction between the two electrons is
\begin{equation}
    H_C(|\boldsymbol{r}_1-\boldsymbol{r}_2|)=\frac{e^2}{4\pi\epsilon_r\epsilon_0|\boldsymbol{r}_1-\boldsymbol{r}_2|} \,,
\end{equation}
where the relative dielectric constant in Si is $\epsilon_r=12.8$. We use the orthonormalized single-electron wave functions to generate the two-electron singlet states with symmetric orbital parts:
\begin{eqnarray}
    &&\Psi_a (\boldsymbol{r}_1)\Psi_a(\boldsymbol{r}_2)\\
    &&\frac{1}{\sqrt{2}}\left[ \Psi_a (\boldsymbol{r}_1)\Psi_b(\boldsymbol{r}_2)+\Psi_a (\boldsymbol{r}_2)\Psi_b(\boldsymbol{r}_1) \right]
\end{eqnarray}
and triplet with antisymmetric orbital wave functions,
\begin{eqnarray}
    \frac{1}{\sqrt{2}}\left[ \Psi_a (\boldsymbol{r}_1)\Psi_b(\boldsymbol{r}_2)-\Psi_a (\boldsymbol{r}_2)\Psi_b(\boldsymbol{r}_1) \right]
\end{eqnarray}
Here $\Psi_a, \Psi_b=\{L_-,L_+,R_-,R_+\}$ are the orthonormalized single electron wave functions. There are 10 singlet and 6 triplet states for two electrons in a double quantum dot.  In this study we do not consider spin-orbit interaction, such that singlet and triplet states are the two-electron eigenstates and their orbital and spin parts are separated.  We will thus not include spin part of the wave functions in any of the following discussions.

In our calculation, we do not include contributions to the Coulomb terms from the expectation values of different bulk valley eigenstates, such as $\langle L_zL_{z}|H_C|L_zL_{-z}\rangle=\langle L_zL_{-z}|H_C|L_{-z}L_{z}\rangle=0$. These expectation values are much smaller than the other terms given in Tables 1 and 2 due to the orthogonal Bloch states in their wave functions. Our final result for S-orbital states has four Coulomb interaction terms.  Their definitions and expectation values are given in Table \ref{tab_s2}.
\begin{center}
\begin{table}[h]
    \centering
    \begin{tabular}{ |p{4cm}||p{6.5cm}|p{3cm}|  }
 \hline
 Coulomb Terms & Coulomb Expression & Values\\
 \hline
 On-site & $u=\langle L_-(\boldsymbol{r}_1)L_-(\boldsymbol{r}_2)|H_C|L_-(\boldsymbol{r}_1)L_-(\boldsymbol{r}_2)\rangle$   & \;\;16.1\;meV \\
 Interdot & $k=\langle L_-(\boldsymbol{r}_1)R_-(\boldsymbol{r}_2)|H_C|L_-(\boldsymbol{r}_1)R_-(\boldsymbol{r}_2)\rangle$  & \;\;\;3.1 \;meV  \\
 Overlap & $s=|\langle L_-(\boldsymbol{r}_1)L_-(\boldsymbol{r}_2)|H_C|L_-(\boldsymbol{r}_1)R_-(\boldsymbol{r}_2)\rangle|$   & \;$-5.7\;\mu$eV\\
 Exchange & $j=|\langle L_-(\boldsymbol{r}_1)R_-(\boldsymbol{r}_2)|H_C|R_-(\boldsymbol{r}_1)L_-(\boldsymbol{r}_2)\rangle|$ & \;\;46.3\;neV \\
 \hline
\end{tabular}
    \caption{Coulomb energy of two electrons in a double quantum dot}
    \label{tab_s2}
\end{table}
\end{center}
The on-site and interdot direct Coulomb repulsion terms $u$ and $k$ defined in Table \ref{tab_s2} have no dependence on the valley-orbit phase of either dot. Whereas the Coulomb overlap $s$ and exchanges $j$ terms do depend on the valley-orbit phases of the quantum dots. The Coulomb overlap term adds to the tunnel coupling of the matrix elements of a single electron with the same phase dependence. For example, the magnitude of tunnel coupling for an electron in the presence of another is modified to $t=t_0+s=-22.8$ $\mu$eV in our particular DQD configuration.
%
%

The Hamiltonian matrix can be divided into blocks depending on the position of the electrons in the DQD: double occupied blocks (both electrons in the left or right dot) and one electron in each dot block: 
\begin{equation}
    H^\alpha=
   \begin{pmatrix}
    H^\alpha_{LL} & H^{\alpha*}_J & T^{\alpha*}_{LR} \\
    H^\alpha_J & H^\alpha_{RR}  & T^{\alpha *}_{RL}\\
      T^\alpha_{LR} & T^\alpha_{RL} &H^\alpha_{LR} \\
    \end{pmatrix}
\end{equation}
where $\alpha=\{s,t\}$, with $s$ and $t$ referring to the singlet and triplet matrices, respectively. Here $H^\alpha_{LL}$ and $H^\alpha_{RR}$ represent the Hamiltonian of the electrons when both are located in either left or the right quantum dots, while $H^\alpha_{LR}$ is for the case of one electron in each dot.The tunnel coupling matrices $T^\alpha_{LR}$ couple the double occupied states to the evenly distributed states. The matrix $H^\alpha_J$ couples the left and right doubly occupied states.

In the main text we calculate the energy of the ground singlet and triplet states by exact diagonalization of the matrix $H^\alpha$.  Qualitatively, considering how large the on-site Coulomb interaction $u$ is, their contribution to the ground states should be quite small.  To help understand the numerical results, here we perform a first-order Schrieffer-Wolff transformation of the Hamiltonian.  In this approximation, we focus on the ${LR}$ block, and account for the double occupied blocks perturbatively. 
%

The expressions for each of the relevant block matrices for singlet and triplet states are, for the left doubly occupied states, 
\begin{equation}
    H^s_{LL}=
   \begin{pmatrix}
    2E_{L}-2|\Delta_{L}|+u & 0 & 0 \\
    0 &2E_{L}+2|\Delta_{L}|+u  & 0\\
      0 & 0 & 2E_{L}+u \\
    \end{pmatrix},
\;\;\;\;\;    
H^t_{LL}=
   \begin{pmatrix}
    2E_{L}+u \\
    \end{pmatrix} \,,
\end{equation}
and for the right doubly occupied states,
\begin{equation}
    H^s_{RR}=
   \begin{pmatrix}
    2E_{R}-2|\Delta_{R}|+u & 0 & 0 \\
    0 &2E_{R}+2|\Delta_{R}|+u  & 0\\
      0 & 0 & 2E_{R}+u \\
    \end{pmatrix},
\;\;\;\;\;    
H^t_{RR}=
   \begin{pmatrix}
    2E_{R}+u \\
    \end{pmatrix} \,.
\end{equation}
The tunnel coupling matrices are,
\begin{equation}
\nonumber
    T^s_{LR}=
   \begin{pmatrix}
    \sqrt{2}t_1 & 0 & t_2  \\
    \sqrt{2}t_2 & 0 & t_1 \\
      0 & \sqrt{2}t_2 & t_1 \\
        0 & \sqrt{2}t_1 & t_2 \\
    \end{pmatrix},
    \;\;\;
       T^t_{LR}=
   \begin{pmatrix}
     \;\;t_2 \\
     \;\;t_1 \\
      \;\;t_1 \\
        -t_2 \\
    \end{pmatrix} \,.
\end{equation}
The one-electron-per-dot block matrices ($LR$) are,
\begin{equation}
\nonumber
    H^s_{LR}=E_L+E_R+k+
   \begin{pmatrix}
    -|\Delta_L|-|\Delta_R|-j\cos^2\varphi & 
-\iota j\sin\varphi \cos\varphi & \iota j\sin\varphi \cos\varphi & j\sin^2\varphi \\
    \iota j\sin\varphi \cos\varphi & -|\Delta_L|+|\Delta_R|+ j\sin^2\varphi  & -j\cos^2\varphi & \iota j\sin\varphi \cos\varphi\\
      -\iota j\sin\varphi \cos\varphi & -j\cos^2\varphi & |\Delta_L|-|\Delta_R|+ j\sin^2\varphi & -\iota j\sin\varphi \cos\varphi \\
      j\sin^2\varphi & -\iota j\sin\varphi \cos\varphi & \iota j\sin\varphi \cos\varphi &  |\Delta_L|+|\Delta_R| -j\cos^2\varphi \\
    \end{pmatrix},
\end{equation}
and
\begin{equation}
\nonumber
    H^t_{LR}=E_L+E_R+k+
   \begin{pmatrix}
    -|\Delta_L|-|\Delta_R|+j\cos^2\varphi & 
\iota j\sin\varphi \cos\varphi & -\iota j\sin\varphi \cos\varphi & -j\sin^2\varphi \\
    -\iota j\sin\varphi \cos\varphi & -|\Delta_L|+|\Delta_R|- j\sin^2\varphi  & j\cos^2\varphi & -\iota j\sin\varphi \cos\varphi\\
      \iota j\sin\varphi \cos\varphi & j\cos^2\varphi & |\Delta_L|-|\Delta_R|- j\sin^2\varphi & \iota j\sin\varphi \cos\varphi \\
      -j\sin^2\varphi & \iota j\sin\varphi \cos\varphi & -\iota j\sin\varphi \cos\varphi &  |\Delta_L|+|\Delta_R| +j\cos^2\varphi \\
    \end{pmatrix}.
\end{equation}
Here $\varphi=\frac{\phi_L-\phi_R}{2}$ is the interdot phase difference. 

There is a natural hierarchy of order of magnitude for the various energy terms: the onsite and interdot Coulomb term $u$ and $k$ are 16 and 3 meV, respectively; the valley splittings and interdot tunnel coupling are in the order of 10 to 100 $\mu$eV; and the exchange term is sub $\mu$eV.  We thus perform a Schrieffer-Wolff transformation and obtain the first order correction to the $LR$ singlet and triplet matrices as
\begin{equation}
    H^{s(1)}_{LR}=
   -\frac{1}{u-k}\begin{pmatrix}
    4|t_1|^2+2|t_2|^2 & 
2t^*_1t_2 & 2t_1t^*_2 & 2|t_2|^2 \\
   2t^*_1t_2 &  2|t_1|^2+4|t_2|^2 &2|t_1|^2 & 2t^*_1t_2 \\
      2t^*_1t_2 & 2|t_1|^2 &   2|t_1|^2+4|t_2|^2 & 2t^*_1t_2\\
      2|t_2|^2 & 2t_1t^*_2 & 2t_1t^*_2 &   4|t_1|^2+2|t_2|^2 \\
    \end{pmatrix},
\end{equation}
and,
\begin{equation}
    H^{t(1)}_{LR}=
   -\frac{1}{u-k}\begin{pmatrix}
    2|t_1|^2+4|t_2|^2 & 
2t^*_1t_2 & 2t_1t^*_2 & -2|t_1|^2 \\
   2t^*_1t_2 &  4|t_1|^2+2|t_2|^2 &-2|t_1|^2 & 2t^*_1t_2 \\
      2t^*_1t_2 & -2|t_1|^2 &   4|t_1|^2+2|t_2|^2 & 2t^*_1t_2\\
      -2|t_2|^2 & 2t_1t^*_2 & 2t_1t^*_2 &   2|t_1|^2+4|t_2|^2 \\
    \end{pmatrix},
\end{equation}
where we use $t_1=t_{--}=t_{++}$ and $t_2=t_{+-}=t_{-+}$ to save some space. The lowest energy levels of the matrices $H^s_{LS}+H^{s(1)}_{LS}$ and $H^t_{LS}+H^{t(1)}_{LS}$ correspond to the ground singlet and triplet states.
Keeping the lowest order terms (i.e. the diagonal matrix elements, considering that $|\Delta_L|, |\Delta_R|\sim 100 \mu $eV), we obtain the ground singlet and triple energy as,
\begin{eqnarray}
     E^s_0&\sim& E_L+E_R-|\Delta_L|-|\Delta_R|+k-\frac{4|t_{--}|^2}{u-k}-\frac{2|t_{-+}|^2}{u-k}+j\cos^2\varphi \,,\\
    E^t_0 &\sim& E_L+E_R-|\Delta_L|-|\Delta_R|+k-\frac{2|t_{-+}|^2}{u-k}-j\cos^2\varphi \,.
\end{eqnarray}
Thus the exchange energy is,
\begin{eqnarray}
\nonumber    E_J &\sim& E^t_0+E^s_0\\
    &=&\frac{4|t_1|^2}{u-k}-2j\cos^2\varphi\;=\left(\frac{4|t|^2}{u-k}-2j\right)\cos^2\varphi\;= E^0_J \cos^2\varphi \,.
\end{eqnarray}
Therefore, within a very good approximation the exchange energy depends on the interdot valley phase difference as $\cos^2\varphi$, agreeing with Fig.~1 of the main text.

\section{Effects of Valley Splittings on Exchange Coupling}
In most experiments related to spin qubits in Si, the magnitude of the valley orbit coupling is much larger than the exchange coupling. However, on some occasions, the magnitude is sufficiently small that it can directly influence the coupled spin dynamics. Here we explore this regime by reducing the magnitude of the valley orbit coupling of the right quantum dot. The valley splitting of the left dot is fixed at 100 $\mu$eV. Figure \ref{fig_s2b} shows the results of exchange splitting for three different values of $|\Delta_R|$. We explain these plots using our analytical expressions from Sec.~\ref{S2}. 

For a smooth interface with $\varphi=0$, the lowest two singlet and triplet states have the energies
\begin{eqnarray}
     E^s_0&=E_{g,s},\;\;\;\;\;\;\;\;\;\;\;&E^s_1=E_{g,s}+2|\Delta_R|+\frac{E_J^0}{2} \,,\\
    E^t_0 &=E_{g,s}+E_J^0,\;\;\;\;\;\;&    E^t_1 =E_{g,s}+2|\Delta_R|+\frac{E_J^0}{2} \,.
\end{eqnarray}
 where $E_{g,s}\sim E_L+E_R-|\Delta_L|-|\Delta_R|+k-\frac{4|t|^2}{u-k}+j $. Note that when $4\Delta_R=E_J$, an accidental degeneracy occurs:  $E_1^s=E_0^t=E_1^t$, as can be seen in panel (b) of Fig.~\ref{fig_s2b}.  When the interdot valley phase difference is $\phi_L-\phi_R=\pi$, the energies of the lowest two singlets and triplets are,
 \begin{eqnarray}
     E^s_0&=E_{g,s}+\frac{E_J^0}{2},\;\;\;\;\;\;\;\;\;\;\;&E^s_1=E_{g,s}+2|\Delta_R| \,,\\
    E^t_0 &=E_{g,s}+\frac{E_J^0}{2},\;\;\;\;\;\;&    E^t_1 =E_{g,s}+2|\Delta_R|+E_J^0 \,.
\end{eqnarray}
Here again an accidental degeneracy occurs at $4\Delta_R=E_J$, when the ground and first excited singlets share the same energy with the ground triplet.

\begin{figure}[h]
\minipage{0.90\textwidth}
  \includegraphics[width=15.5cm, height= 5.0cm]{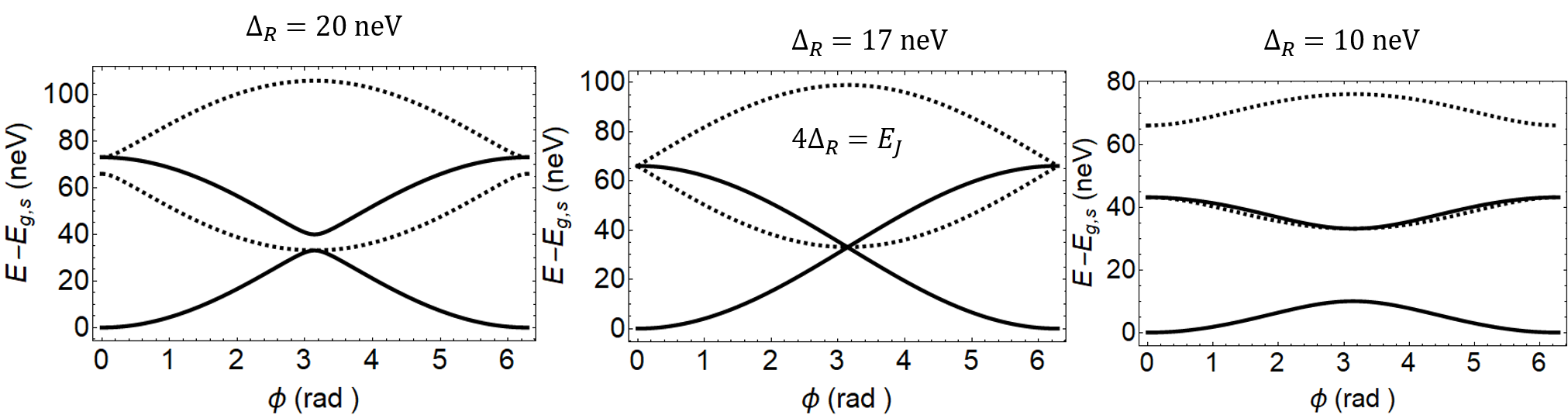}
\endminipage\hfill
   \caption{Change in the energy levels of the lowest two ground singlet and triplet states with $\phi$. We use three different values of the magnitudes of the valley orbital coupling in the right quantum dot.
}\label{fig_s2b}
\end{figure}

As the valley splitting of the right quantum dot decreases, valley phase $\phi_R$ gradually becomes ill defined and irrelevant, when the valley phase in the left dot is the dominant factor in determining the state composition in the DQD. 

In a hypothetical situation where we consider the valley splittings in both dots are zero, the four lowest singlet and triplet energies in the diagonals have the eigen values
\begin{eqnarray}
    E_s &\sim &E_L+E_R+k+\left\{-\frac{4|t|^2}{u-k}+j,-\frac{4|t|^2}{u-k}+j,-\frac{4|t|^2}{u-k}+j,-j\right\}\\
    E_t &\sim &E_L+E_R+k+\left\{-\frac{4|t|^2}{u-k}+j,-j,-j,-j\right\}\\.
\end{eqnarray}
Thus the ground state is four-fold degenerate, containing three singlet states and one triplet state. The excited energy level is also four-fold degenerate, containing three triplets and one singlet. With our dots configuration, the energy gap between the singlet and triplet energy level is $\frac{4|t|^2}{u-k}-2j$. 

\section{Effect of the step position on valley orbit couplings }

Interface roughness is intrinsically present at the Si-barrier interface during the growth process. An example of such roughness is interface steps in a Si-SiGe heterostructure.  Our previous results show that a single step can reduce the valley splittings by more than 70 $\%$ and a nearly $\pi$ shift in the valley phase in the S orbital state. Our aim here is to examine the effects of an interface step on the exchange coupling in a Si DQD, especially in the presence of higher-energy orbital states.

Consider when we include the three lowest harmonic oscillator energy levels, i.e. the s-, p-, and d-levels.  There are thus six orbital states.  When valley is introduced, there is a valley-orbit coupling term for each of these orbitals, so that there are six diagonal valley orbit coupling terms. In Fig.~\ref{fig_s4} (a) and \ref{fig_s4} (b), we show the step-position dependence of the valley splittings and phases in the diagonal elements of the left quantum dot with a monolayer step.  Note that the influence of the step on $\Delta_{p_yp_y}$ and $\Delta_{d_{yy}d_{yy}}$ are the same as on $\Delta_{ss}$, and also $\Delta_{p_xp_x}=\Delta_{d_{xy}d_{xy}}$.  Similar to the ground orbital state, the valley splittings for the other orbital states are also reduced when the step cuts through the quantum dot, reaching their minima when the step is located at the center of the dot. Furthermore, the valley orbit coupling behavior is not the same for different orbital states due to the shape of the Fock-Darwin wave functions along the $x$ direction.

\begin{figure}[h]
\minipage{0.85\textwidth}
  \includegraphics[width=13cm, height= 5.0cm]{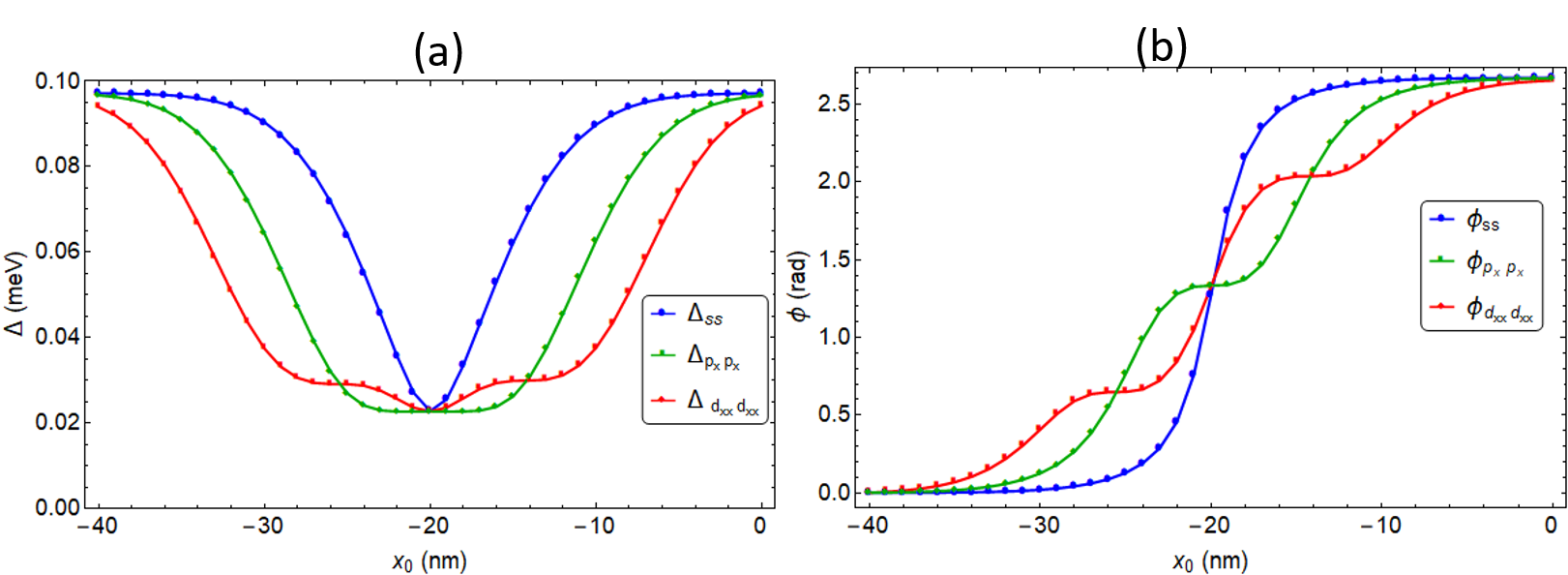}
\endminipage\hfill
   \caption{Changes in the magnitudes (a) and phases (b) in the valley orbital coupling in the diagonal elements as a function of step position.
}\label{fig_s4}
\end{figure}

Valley phases for each orbital state also depend on the step location, as shown in Fig.~\ref{fig_s4}(b). The range of the valley phase is from 0 to 0.85$\pi$. The change in the valley phase of $\phi_{d_{xx}d_{xx}}$ is wider with two stairs like features, because of the presence of three nodes in the $x$ direction in the $d_{xx}$ orbital. On the other hand, $\phi_{p_{x}p_{x}}=\phi_{d_{xy}d_{xy}}$, with one node present in their wave functions along the $x$ direction. Lastly, $\phi_{ss}$ changes the most sharply as the s-orbital is the smallest.

\begin{figure}[h]
\minipage{0.85\textwidth}
  \includegraphics[width=7cm, height= 5.0cm]{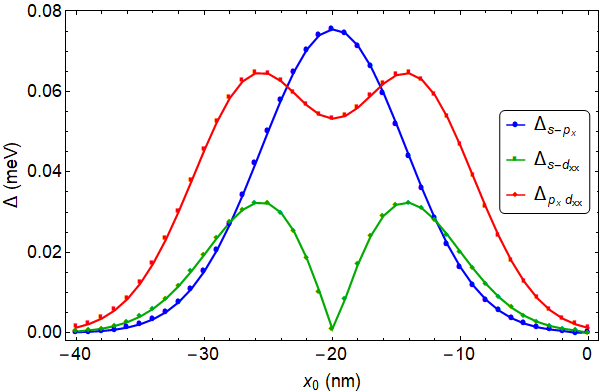}
\endminipage\hfill
   \caption{Change in the magnitude on off-diagonal terms of the valley orbit coupling with the step location.
}\label{fig_s5}
\end{figure}

With the presence of higher orbitals, new off-diagonal valley orbit coupling terms also appear. In this paper, we consider up to the d-orbital states in our calculations, and the step orientation is along the $y$ direction so that its position can be defined by its $x$ coordinate. Under these conditions there are three non-vanishing off-diagonal terms, $\Delta_{sp_x}$, $\Delta_{p_xd_{xx}}$ and $\Delta_{sd_{xx}}$. In Fig.~\ref{fig_s5} we show how the position of the step affects the magnitude of each of these off-diagonal terms.  In particular, the magnitude of $\Delta_{sp_x}$ reaches a maximum value of $\sim 0.075$ meV when the step is at the center of the dot. $\Delta_{p_xd_{xx}}$ and $\Delta_{sd_{xx}}$ are also finite. The phases of these off-diagonal terms are not affected much and hence have not been shown here. In short, the presence of these finite off-diagonal elements in the valley orbit coupling matrix is quite important in our calculation of the exchange splitting.

\section{Effects of the Higher Valley Orbit Couplings on Exchange Interaction}

In this section, we show the changes in the results of the exchange energy due to the valley-orbit coupling of the ground and higher orbital states. We have included up to d-orbital levels in our calculations.  To single out the effect of each term, we change one valley-related parameter at a time and keep all other variables at the same values for a smooth interface. In Fig.~\ref{D_EX}, we show effects of the valley phases $\phi_{ss}$ and $\phi_{p_xp_x}$ on the exchange energy. The other phases have essentially no impact on the exchange interaction.

\begin{figure}[h]
\minipage{0.4\textwidth}
  \includegraphics[width=7cm, height= 4.8cm]{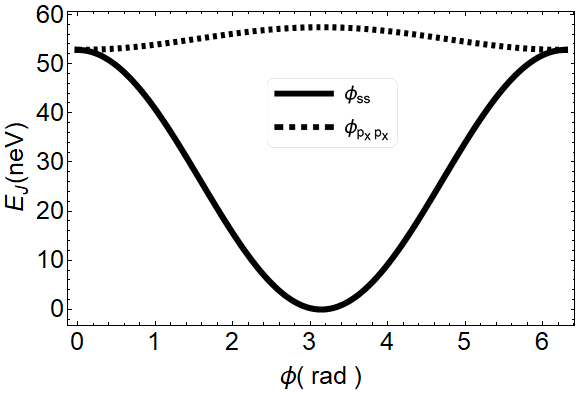}
\endminipage\hfill
   \caption{Effect of the valley phases $\phi_{ss}$ and $\phi_{p_xp_x}$ on the exchange interaction.
}\label{D_EX}
\end{figure}

In Fig.~\ref{D_EX} the effect of the valley phase $\phi_{ss}$ on the exchange interaction is qualitatively the same as in Fig. 1.  The exchange splitting has a maximum value at $\phi_{ss}=0$ and reaches a minimum when $\phi_{ss}=\pi$. The behavior of the plot is also proportional to $\cos^2\varphi$. The difference here is that the higher orbital states modify the value of the exchange energy for the smooth interface at $\phi_{ss}=0$. 

The exchange energy has the opposite behavior, albeit in much smaller magnitude, when we change the valley phase $\phi_{p_xp_x}$, reaching a maximum value when $\phi_{p_xp_x}=\pi$. At $\phi_{p_xp_x}=\pi$ the two valley states $L_{-,p_x}$ and $L_{+,p_x}$ switch their order in energy. This swapping of the energy levels leads to a shift of the energy levels by $\Delta$ in the diagonal elements.  However, both these orbitals have the same couplings with the s- and d-orbitals. Hence we only see a slight change in the singlet and triplet energy spectrum. The amount of change in the exchange energy is directly proportional to the magnitude of the valley-orbit couplings in the $p$ state.

\begin{figure}[h]
\minipage{0.4\textwidth}
  \includegraphics[width=7cm, height= 4.8cm]{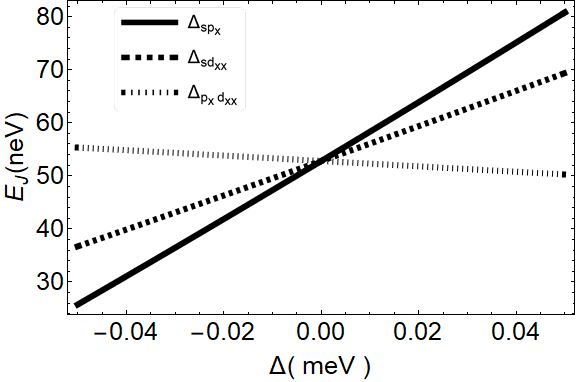}
\endminipage\hfill
   \caption{Effect of the magnitudes of the off diagonal elements on the exchange interaction. 
}\label{Off_Ex}
\end{figure}

In Fig.~\ref{Off_Ex} we show the effects of the off-diagonal valley-orbit couplings ($\Delta_{sp_x}, \Delta_{sd_{xx}}$ and $\Delta_{p_xd_{xx}}$) on the exchange splitting $E_J$.  Clearly, $E_J$ changes linearly here. The slope of each line is different, and can be positive or negative. The most important contribution is from $\Delta_{sp_x}$. Exchange energy rises by up to 50\% of its initial value as $\Delta_{sp_x}$ increases from 0 to $0.05$ meV. 
Here we have changed the valley parameters one by one while keeping all other variables the same as in the smooth interface. If an interface step is introduced, all of these variables will be changing with the step position simultaneously. Nevertheless, the results obtained here help us to identify the most important factor ($\Delta_{sp_x}$) and the qualitative behavior of the relevant dependence (linear).

\end{widetext}

\bibliography{EX}

\end{document}